%% file: elsarticle-template-harv.tex
\documentclass[preprint,12pt,authoryear]{elsarticle}

\usepackage{amssymb,amsmath}
\usepackage{multirow}
\usepackage{amsthm}
\usepackage{booktabs}
\usepackage{multirow}

\usepackage{here}
\usepackage[hyphens]{url}
\usepackage{subfig}
\usepackage{bm}

\journal{Spatial Statistics}

\begin{document}

\begin{frontmatter}



\title{Modeling heterogeneity in higher-order moments \\ while preserving mean and variance: \\ application to spatio-temporal modeling}


\author[label1,label2]{Hajime Kuno\corref{cor1}}
\ead{hakuno103@gmail.com}
\cortext[cor1]{Corresponding author}
\author[label3]{Daisuke Murakami}

\affiliation[label1]{
    organization={The Graduate University for Advanced Studies},
    addressline={Shonan Village}, 
    city={Hayama},
    postcode={240-0193}, 
    state={Kanagawa},
    country={Japan}
}

\affiliation[label2]{
    organization={BrainPad inc.},
    addressline={3-1-1 Roppongi}, 
    city={Minato-ku},
    postcode={108-0071}, 
    state={Tokyo},
    country={Japan}
}

\affiliation[label3]{
    organization={Institute of Statistical Mathematics},
    addressline={10-3 Midori-cho}, 
    city={Tachikawa},
    postcode={190-8562}, 
    state={Tokyo},
    country={Japan}
}
\begin{abstract}
\input{abstract}
\end{abstract}


\end{frontmatter}

\input{1_introduction}

\input{2_CSN_and_CSNS_distribution}

\input{3_proposed_method} 

\input{4_application_to_spatiotemporal_data}

\input{5_bayesian_inference}

\input{6_simulation_studies}

\input{7_application}

\input{8_conclusion}

\section*{Acknowledgments}
This work was supported by JSPS KAKENHI Grant Number 24K00175.

\bibliographystyle{elsarticle-harv}
\bibliography{refs}

\end{document}

%% file: abstract.tex
In this study, we propose a general model capable of addressing heterogeneity in higher-order moments while preserving mean and variance, including the t, Laplace, and skew-normal distributions as special cases. Our model flexibly accommodates variations in tail heaviness and asymmetry at each data point while maintaining interpretability similar to normal distribution models. Notably, it is closed under linear transformations and provides explicit analytical expressions for skewness and kurtosis. The proposed model is applied to spatial and temporal data analysis, demonstrating that its properties vary based on the chosen matrix decomposition approach. To facilitate efficient inference, we develop a Bayesian estimation method using data augmentation, which is particularly effective for temporal models. Simulation studies confirm that accounting for heterogeneity in higher-order moments enhances parameter estimation accuracy and predictive performance. To illustrate real-world applicability, we analyze production functions across U.S. states. The results indicate that our model effectively captures heterogeneity in higher-order moments, leading to superior model fit in empirical data analysis. 

%% file: 1_introduction.tex
\section{Introduction}
The modeling of higher-order moments, such as skewness and kurtosis, is crucial in many real-world applications \citep{tagle2019non}. For example, infection risk may exhibit a skewed distribution with extremely high values during epidemic periods, and the strength of skewness can vary across different regions and time periods. Therefore, describing the process using a non-Gaussian model that can flexibly handle such heterogeneous skewness and higher-order moments is desirable.

To account for non-Gaussianity, a wide variety of methods have been developed, as reviewed by \cite{yan2020multivariate}. A prominent example is the closed-skew normal (CSN) distribution, known for its expressive power \citep{azzalini1985class, azzalini1999statistical}. However, replacing the normal distribution in a model with a CSN distribution often disrupts the mean and variance properties, which are essential for interpretability. In particular, spatio-temporal models—actively studied in fields such as disease mapping \citep{best2005comparison}, species distribution modeling \citep{dormann2007methods}, and socio-economic analysis \citep{anselin2022spatial} —rely on the mean regression function to interpret the influence of the specified covariates and covariance specified to quantify the strength of spatial and temporal correlations.

\cite{marquez2022flexible} addressed this issue by proposing a flexible subclass of the CSN distribution that maintains the mean and variance. This property enhances the interpretability of parameters such that when the mean is specified by a regression function, the coefficients and variance parameters can be interpreted similarly to those in a conventional Gaussian regression model. Additionally, \cite{linda2024variational} proposed an extended model known as the closed skew normal subclass (CSNS) distribution, which addresses the heterogeneity of higher-order moments. The CSNS distribution still has limitations in the range of achievable skewness and kurtosis, restricting its representational flexibility. Moreover, \cite{linda2024variational} used the CSNS distribution as an approximation in variational inference, rather than as a component of modeling. Notably, although the model's properties vary depending on the matrix decomposition method used internally, it is restricted to LU decomposition. Similarly, no established Bayesian estimation method is available despite its potential merits for modeling uncertainty in spatio-temporal phenomena \citep{cressie2011statistics,marquez2022flexible}.

To overcome these limitations, we propose a general model that encompasses the CSNS distribution as a special case and accommodates the heterogeneity of higher-order moments while preserving the mean and variance. Subsequently, several properties of the proposed model are presented. In particular, since the characteristics of the model vary depending on the matrix factorization method employed, we explain the specific matrix factorization techniques suitable for adapting to spatio-temporal data. Furthermore, we develop an efficient Bayesian inference method that is particularly effective for temporal data.

The remainder of this paper is organized as follows. Section 2 provides background on the CSN and CSNS distributions. Section 3 introduces the proposed model and its theoretical properties. Section 4 investigates the impact of different matrix factorization methods, particularly in spatio-temporal context. Section 5 develops an efficient Bayesian inference method. Section 6 presents the simulations conducted to the model estimation accuracy, and Section 7 applies the model to the estimation of production functions across U.S. states.

%% file: 2_CSN_and_CSNS_distribution.tex
\section{CSN and CSNS distribution}

The CSN distribution is an extension of the multivariate normal distribution that incorporates higher-order moments. For \(\bm{s} \sim \mathcal{CSN}_{N, M}(\bm{\mu}, \bm{\Sigma}, \bm{D}, \bm{\nu}, \bm{\Omega})\), the probability density function is given by:
\begin{align*}
p(\bm{s})=\phi_N(\bm{s} ; \bm{\mu}, \bm{\Sigma}) \frac{\Phi_M(\bm{D}(\bm{s}-\bm{\mu}) ; \bm{\nu}, \bm{\Omega})}{\Phi_M\left(\bm{0}_M ; \bm{\nu}, \bm{\Omega} + \bm{D} \bm{\Sigma} \bm{D}^\top\right)}.
\end{align*}
Here, \(\bm{\mu} \in \mathbb{R}^p\), \(\bm{\nu} \in \mathbb{R}^q, \bm{D}\in\mathbb{R}^{q\times p}\), and \(\bm{\Sigma} \in \mathbb{R}^{N\times N}, \bm{\Omega} \in \mathbb{R}^{M\times M}\) are covariance matrices. \(\phi_N(\bm{s}; \bm{\mu}, \bm{\Sigma})\) and \(\Phi_N(\bm{s}; \bm{\mu}, \bm{\Sigma})\) denotes the probability distribution function and cumulative distribution function of an \(N\)-dimensional normal distribution with mean \(\bm{\mu}\) and covariance \(\bm{\Sigma}\), respectively. Finally, \(\bm{0}_M\) represents an \(M\)-dimensional zero vector.

As a modified version of the CSN distribution, the CSNS distribution ensures that the mean and variance remain consistent with those of a multivariate normal distribution. The CSNS distribution is derived by first defining the following random variable:
\begin{align*}
    \theta_i u_i + v_i, u_i \sim \mathcal{TN}(0, 1), v_i \sim \mathcal{N}_1(0, 1),
\end{align*}
where \(\mathcal{N}_N(\bm{\mu}, \bm{\Sigma})\) denotes a multivariate normal distribution with mean \(\bm{\mu}\) and covariance \(\bm{\Sigma}\), and \(\mathcal{TN}(\mu, \sigma^2)\) denotes a normal distribution \(\mathcal{N}_1(\mu, \sigma^2)\) truncated below \(0\). The mean of the above random variable is \(\sqrt{2 / \pi}\theta_i \) and the variance is \(f_i^{-2} = (\pi+(\pi-2)\theta_i^2) / \pi\). To standardize the variable, we define \(t_i = f_i (\theta_i (u_i - \sqrt{2/\pi}) + v_i)\), ensuring that \(t_i\) has a mean of \(0\) and a variance of \(1\). Setting
\begin{align}
    \bm{s} = \bm{\mu} + \bm{\Sigma}^{LU} \bm{t}, \label{CSNS}
\end{align}
guarantees that \(\bm{s}\) maintains mean \(\bm{\mu}\) and covariance \(\bm{\Sigma}\). Here, \(\bm{s} = (s_1, \ldots, s_N)^\top\), \(\bm{t} = (t_1, \ldots, t_N)^\top\), and \(\bm{\Sigma}^{LU}\) is the LU decomposition of \(\bm{\Sigma}\). The distribution of \(\bm{s}\) can be expressed as the CSN distribution:
\begin{align*}
\mathcal{CSN}_{N, N}&(\bm{\mu} - \bm{\Sigma}^{LU} \bm{F} \bm{\mu}_{\bm{u}}, \bm{\Sigma}^{LU} \bm{F} (\bm{I}_N + \bm{\Theta} \bm{\Theta}) \bm{F} (\bm{\Sigma}^{LU})^\top,  \\
&\qquad \bm{\Theta} \bm{F} (\bm{\Sigma}^{LU}\bm{F}(\bm{I}_N + \bm{\Theta} \bm{\Theta})^{1/2})^{-1}, \bm{0}_N, \bm{I}_N - \bm{\Theta}\bm{F} \bm{F}\bm{\Theta}),
\end{align*}
where \(\bm{\mu}_{\bm{u}} = (\sqrt{2 / \pi}\theta_1, \ldots, \sqrt{2 / \pi}\theta_N)^\top\), \(\bm{F} = diag(f_1, \ldots, f_N)\), \(\bm{\Theta} = diag(\theta_1, \ldots, \theta_N)\), and \(\bm{I}_N\) is the identity matrix.

\begin{figure}[ht]
  \centering
  \includegraphics[width=0.8\linewidth]{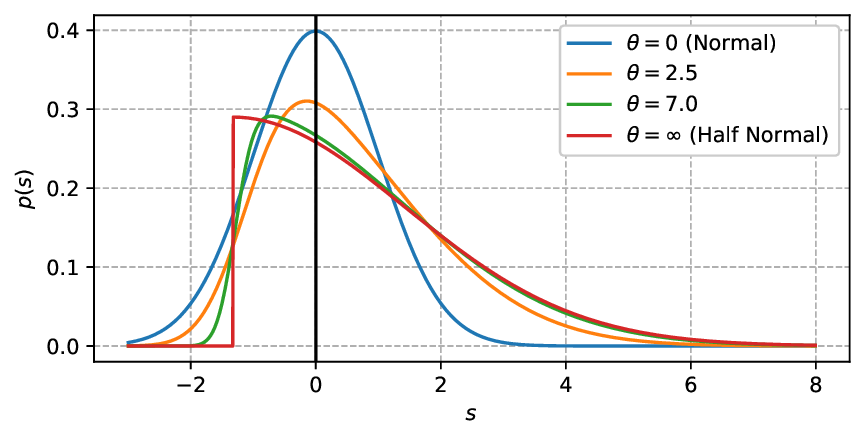}
  \caption{Probability density function of the CSNS distribution for various values of \(\theta\).}
  \label{fig:pdf}
\end{figure}

Figure \ref{fig:pdf} illustrates the probability density function of the CSNS distribution with mean \(0\) and variance \(1\) for different values of \(\theta\). When \(\theta = 0\), the CSNS distribution coincides with the normal distribution. As \(\theta\) increases, the skewness increases, and the distribution approaches a shifted half-normal distribution.

%% file: 3_proposed_method.tex
\section{Proposed method}
In this section, a general model that can handle heterogeneity in higher-order moments while preserving the mean and variance is proposed, based on the idea of the CSNS distribution. In Section 3.1, we provide an overview of the model along with several illustrative examples. Subsequently, in Section 3.2, we discuss the model's closure under linear transformations as well as skewness and kurtosis.

\subsection{A general model for heterogeneous highe-order moments}
To derive the proposed model, we first consider an \(N\)-dimensional random variable  \(\bm{s} = (s_1, \ldots, s_N)^\top\) defined as:
\begin{align}
    \bm{s} = \bm{\mu} + \bm{\Sigma}^{1/2}\bm{t}, \label{proposed model}
\end{align}
where \(\bm{t}\) is a random variable with independent components of mean \(0\) and variance \(1\), and \(\bm{\Sigma}^{1/2}\) is any matrix satisfying \(\bm{\Sigma}^{1/2}(\bm{\Sigma}^{1/2})^\top = \bm{\Sigma}\). The mean and variance of \(\bm{s}\) are \(\bm{\mu}\) and \(\bm{\Sigma}\), respectively. As observed in Equation \eqref{CSNS}, the original CSNS distribution considers the LU decomposition. However, since the mean and variance can be preserved regardless of the choice of matrix decomposition, we employ a more general matrix decomposition in the definitions to facilitate subsequent discussions. 

There are many possible ways to design \( \mathbf{t}=(t_1,\ldots,t_N)^T \). In this study, we consider a random variable that can be expressed as follows:
\begin{align*}
    t_i = f(\bm{\theta}_i) (g(\bm{u}_i, \bm{\theta}_i) + h(\bm{u}_i, \bm{\theta}_i)v_i), v_i \sim \mathcal{N}(0, 1),
\end{align*}
where \(\bm{u}_i\) is a \(d_1\) dimensional random variable independent of \(v_i\), and \(\bm{\theta}_i\) is a \(d_2\) dimensional parameter. The function \( f : \mathbb{R}^{d_2} \rightarrow \mathbb{R}\) is the reciprocal of the standard deviation of \( g(\bm{u}_i, \bm{\theta}_i) + h(\bm{u}_i, \bm{\theta}_i)v_i \) that ensures standardization. \( g  : \mathbb{R}^{d_1} \times\mathbb{R}^{d_2} \rightarrow \mathbb{R} \) is a random variable with mean \(0\). \(h : \mathbb{R}^{d_1} \times\mathbb{R}^{d_2} \rightarrow \mathbb{R}\) is chosen so that the variance of \(g(\bm{u}_i, \bm{\theta}_i) + h(\bm{u}_i, \bm{\theta}_i)v_i \) is positive. We adopted the following notations \( f_i = f(\bm{\theta}_i) \), \( g_i = g(\bm{u}_i, \bm{\theta}_i) \), and \( h_i = h(\bm{u}_i, \bm{\theta}_i) \).

Table \ref{tab:examples of the proposed model} summarizes the specific instances of the proposed model. Here, \(\mathcal{IG}(a, b)\) denotes an inverse gamma distribution with parameters \(a\) and \(b\), \(\mathcal{E}(\mu)\) denotes an exponential distribution with mean \(\mu\), and \(\mathcal{IN}(\mu, \lambda)\) denotes an inverse Gaussian distribution with parameters \(\mu\) and \(\lambda\).

\begin{table}[!hbt]
\centering
\begin{tabular}{lcccc}
\hline
& \(f_i\) & \(g_i\) & \(h_i\) & \(p(\bm{u}_i)\)  \\ \hline
t & \(\sqrt{\frac{\theta_i-2}{\theta_i}}\) & \(0\) & \(\sqrt{u_i}\) & \(\mathcal{IG}(\theta_i/2, \theta_i/2)\) \\
Laplace & \(1\) & \(0\) & \(\sqrt{u_i}\) & \(\mathcal{E}(1)\) \\
NIG & \(\sqrt{\frac{\theta_{i2}}{\theta_{i1}^2+\theta_{i2}}}\) & \(\theta_{i1}(u_i-1)\) & \(\sqrt{u_i}\) & \(\mathcal{IN}(1, \theta_{i2})\) \\
AL & \(\frac{1}{\sqrt{\theta_i^2+1}}\) & \(\theta_i (u_i-1)\) & \(\sqrt{u_i}\) & \(\mathcal{E}(1)\) \\
SN & \(\sqrt{\frac{\pi}{\pi+(\pi-2)\theta_i^2}}\) & \(\theta_i(u_i-\sqrt{\frac{2}{\pi}})\) & \(1\) & \(\mathcal{TN}(0, 1)\) \\
ST & \(\sqrt{\frac{\pi(\theta_{i2}-2)}{\theta_{i2}(\pi+(\pi-2)\theta_{i1}^2)}}\) & \(\sqrt{u_{i2}}\theta_{i1}(u_{i1}-\sqrt{\frac{2}{\pi}})\) & \(\sqrt{u_{i2}}\) & \begin{tabular}{c}\(u_{i1}\sim\mathcal{TN}(0, 1)\) \\ \(u_{i2} \sim \mathcal{IG}(\frac{\theta_{i2}}{2}, \frac{\theta_{i2}}{2})\)\end{tabular} \\ \hline
\end{tabular}
\caption{Examples of the proposed model.}
\label{tab:examples of the proposed model}
\end{table}

First, a simple example is the t-distribution, which requires \(\theta_i > 2\) for the mean and variance to exist. Similarly, the Laplace distribution can be considered. Both distributions are symmetric with zero skewness. A more generalized class of distributions that allow for non-zero skewness is the family of normal mean-variance mixtures. Examples of such distributions include the normal-inverse Gaussian (NIG) and asymmetric Laplace (AL) distributions. A further generalization of the NIG distribution is the generalized hyperbolic distribution.

The most important example is the skew normal (SN) distribution, because, as discussed in the previous section, \(\bm{s}\) follows the CSNS distribution, which is a subclass of the CSN distribution. Moreover, the CSN distribution and its derivative, the unified skew normal distribution, are known to exhibit conjugacy in linear regression, probit, multinomial probit, and Tobit models, making them an active area of research in recent years \citep{niccolo2023bayesian}. Additionally, as a scale mixture of the SN distribution, the skew-t distribution (ST) emerges as a natural extension. An important property of this model is that it is closed under linear transformations of the SN distribution through scale mixtures \citep{wang2024multivarite}.

Furthermore, since the distributions of each \(t_i\) do not necessarily have to be identical, more general models, such as combinations of multiple distributions, can also be treated within the same framework.

\subsection{Properties of the proposed model}
An important property of this model is that it is closed under linear transformations. Specifically, for any \(N \times N\) matrix \(\bm{A}\), we have \(\bm{A} \bm{s} = \bm{A} \bm{\mu} + \bm{A} \bm{\Sigma}^{1/2} \bm{t}\), indicating that the model is closed under linear transformations.

Furthermore, if the third and fourth moments of \( t_i \) exist, the skewness and kurtosis can be analytically derived as follows;
\begin{align*}
    Skewness(s_i) &= \frac{\sum_{j=1}^N \gamma_j (\Sigma^{1/2}_{i, j})^3}{(\Sigma_{i, i})^{3/2}}\\
    Kurtosis(s_i) &= \frac{\sum_{j=1}^N \delta_j (\Sigma^{1/2}_{i, j})^4}{(\Sigma_{i, i})^2} - 3
\end{align*}
where \(\Sigma_{i, j}\) is the element in the \(i\)th row and \(j\)th column of \(\bm{\Sigma}\), and \(\Sigma^{1/2}_{i, j}\) is the element in the \(i\)th row and \(j\)th column of \(\bm{\Sigma}^{1/2}\); \(\gamma_i\) represents the third moment of \(x_i\); and \(\delta_i\) denotes its fourth moment. Table \ref{tab:third and fourth moment of distributions} presents the third and fourth moments for each model. In the case of the \( t \)-distribution, the third moment exists only if \(\theta_i > 3\), and the fourth moment exists only if \(\theta_i > 4\). Similarly, in the case of the ST distribution, the third moment exists only if \(\theta_{i2} > 3\), and the fourth moment exists only if \(\theta_{i2} > 4\).

\begin{table}[!hbt]
\centering
\begin{tabular}{lcc}
\hline
& Third moment & Fourth moment \\ \hline
t & \(0\) & \(3+\frac{6}{\theta_i-4}\) \\
Laplace & \(0\) & \(6\) \\
NIG & \(\frac{3\theta_{i1}}{\sqrt{\theta_{i2}(\theta_{i1}^2+\theta_{i2})}}\) & \(3+\frac{15}{\theta_{i2}} - \frac{12}{\theta_{i1}^2+\theta_{i2}}\) \\
AL & \(\frac{\theta_i(3+2\theta_i^3)}{(1+\theta_i^2)^{3/2}}\) & \(9 - \frac{3}{(1+\theta_i^2)^2}\) \\
SN & \(\frac{\sqrt{2}(4-\pi)\theta_i^3}{(\pi+(\pi-2)\theta_i^2)^{3/2}}\) & \(3 + \frac{4 \theta_i^2(2(\pi-3)\theta_i^2 - 3 (\pi-2))}{(\pi+(\pi-2)\theta_i^2)^2}\) \\
ST & \(\frac{(4-\pi)}{2}\frac{\Gamma((\theta_{i2}-3)/2)}{\Gamma(\theta_{i2}/2)}\left(\frac{\theta_{i1}\sqrt{\theta_{i2}-3}}{\sqrt{\pi+(\pi-2)\theta_{i1}^2}}\right)^{3}\) & \(\frac{\theta_{i2}-2}{\theta_{i2}-4}\left(3 + \frac{4 \theta_{i1}^2(2(\pi-3)\theta_{i1}^2 - 3 (\pi-2))}{(\pi+(\pi-2)\theta_{i1}^2)^2}\right)\) \\ \hline
\end{tabular}
\caption{Third and fourth moments of distributions.}
\label{tab:third and fourth moment of distributions}
\end{table}

Mardia's skewness and kurtosis can also be analytically derived, and both are independent of the mean and variance. For two random variables \(\bm{s}\) and \(\bm{s}^*\) following the same distribution with mean \(\bm{\mu}\) and covariance matrix \(\bm{\Sigma}\), Mardia's skewness \(\mathrm{MS}(\bm{s})\) and kurtosis \(\mathrm{MK}(\bm{s})\) follow the expressions given below:
\begin{align*}
\mathrm{MS}(\bm{s}) &= \mathbb{E}[((\bm{s}-\bm{\mu})^\top\bm{\Sigma}^{-1}(\bm{s}^*-\bm{\mu}))^3], \\
\mathrm{MK}(\bm{s}) &= \mathbb{E}[((\bm{s}-\bm{\mu})^\top\bm{\Sigma}^{-1}(\bm{s}-\bm{\mu}))^2].
\end{align*}
Since \(\bm{s} = \bm{\mu} + \bm{\Sigma}^{1/2}\bm{t}\),
\begin{align*}
\mathrm{MS}(\bm{s}) &= \mathbb{E}[(\bm{t}^\top\bm{t}^*)^3] = \sum_{i=1}^N \gamma_i^2, \\
\mathrm{MK}(\bm{s}) &= \mathbb{E}[(\bm{t}^\top\bm{t})^2]  = N(N+2) + \sum_{i=1}^N \delta_i.
\end{align*}
These properties imply that the first, second, and higher-order moments can be designed independently.

%% file: 4_application_to_spatiotemporal_data.tex
\section{Application to spatio-temporal modeling}
In the proposed model, characteristics vary based on the matrix decomposition method. Although the scope of this model is not necessarily limited to spatial or temporal data, it is well-suited for explaining changes in model properties. Additionally, since spatial and temporal models are often designed based on the mean and variance, preserving these statistical properties. In Section 4.1, we consider applications to spatial models, followed by applications to temporal models in Section 4.2.

\subsection{Application to spatial modeling}
Many spatial models, such as the spatial autoregressive (SAR) and the conditional autoregressive (CAR) model, capture spatial correlations by assuming that the random variable of interest follows \(N_K(\bm{\mu}, \bm{\Sigma})\) and designing \(\bm{\Sigma}\) accordingly, where \(K\) denotes the number of locations. 

When applying the proposed model to spatial data, we substitute the mean and variance of the spatial model into the corresponding parameters of the proposed model. However, the model's properties can significantly vary depending on the chosen matrix decomposition method. For example, similar to the standard SAR model, consider a model \(\bm{y} = \eta \tilde{\bm{W}} \bm{z} + \bm{s}\). where \(\eta\) is a parameter that determines the strength of the spatial correlation, and \(\tilde{\bm{W}}\) is the row-standardized adjacency matrix. Here, we assume that \(\bm{s}\) follows the proposed model, that is, it satisfies Equation \eqref{proposed model}. In this case,  \(\bm{y} = (\bm{I} - \eta\tilde{\bm{W}})^{-1} \bm{\mu} + (\bm{I} - \eta\tilde{\bm{W}})^{-1}\bm{\Sigma}^{1/2}\bm{t}\) follows the model, because of the property of
closure under linear transformations. This indicates that using \((\bm{I}_K - \eta \tilde{\bm{W}})^{-1}\bm{\Sigma}^{1/2}\) for matrix decomposition is natural, and properties are preserved.

On the other hand, for CAR models, and more generally for data where the column-wise ordering has no inherent meaning, the standard practice is to use the matrix square root. For example, in the Leroux model, which is a type of CAR model, the covariance matrix is specified as follows:
\begin{align}
    \bm{\Sigma} &= \omega^2 \bm{Q}^{-1} \label{Leroux model}\\
    \bm{Q} &= \eta ({\rm diag}(\bm{W}\bm{1}_K)-\bm{W}) + (1-\eta)\bm{I}_K, \notag
\end{align}
where \(\bm{W}\) is the adjacency matrix. The reason for using the matrix square root in the CAR model is that the model constructed using the matrix square root exhibits symmetry with respect to changes in the ordering of the data, whereas the model using the Cholesky decomposition or LU decomposition is not symmetric under permutation. Let the permuted random variable of \(\bm{y}\), which follows the proposed model, be \(\tilde{\bm{y}}=(y_{\chi(1)}, \ldots, y_{\chi(N)})^\top\), where \(\bm{P}\) is the permutation matrix such that \(\bm{P}\bm{y}=\tilde{\bm{y}}\). Consider the randomly permuted variable \(\bm{P} \bm{y} = \bm{P} \bm{\mu} + \bm{P} \bm{\Sigma}^S \bm{x}\) when using the matrix square root \(\bm{\Sigma}^S\). Since the matrix square root of \(\tilde{\bm{\Sigma}} = \bm{P} \bm{\Sigma} \bm{P}^\top\) is given by \(\bm{P} \bm{\Sigma}^S \bm{P}^\top\), we obtain  
\begin{align*}
    \bm{P} \bm{y} = \bm{P} \bm{\mu} + \bm{P} \bm{\Sigma}^S \bm{P}^\top \bm{P} \bm{x} = \tilde{\bm{\mu}} + \tilde{\bm{\Sigma}}^S \tilde{\bm{x}}
\end{align*}
which implies that the distribution of \(\bm{P} \bm{y}\) is identical to that of \(\tilde{\bm{y}}\). On the other hand, for general matrix decompositions such as Cholesky decomposition or LU decomposition, \(\tilde{\bm{\Sigma}}^{1/2}\) does not satisfy \(\tilde{\bm{\Sigma}}^{1/2} = \bm{P} \bm{\Sigma}^{1/2} \bm{P}^\top\), leading to the loss of symmetry with respect to permutation.

\begin{figure}[!hbt]
    \centering
    \subfloat[Skewness of CSNS distribution using the Cholesky decomposition.]{%
        \includegraphics[width=0.9\textwidth]{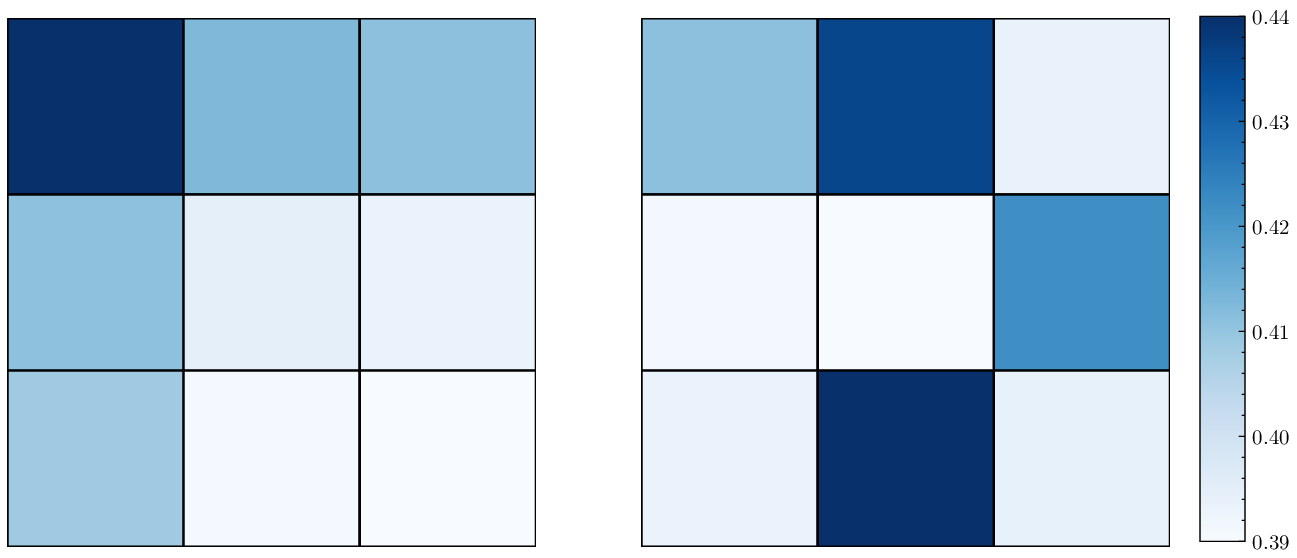}
        \label{fig:subfig1}
    }\\
    \subfloat[Skewness of CSNS distribution using the matrix square root.]{%
        \includegraphics[width=0.9\textwidth]{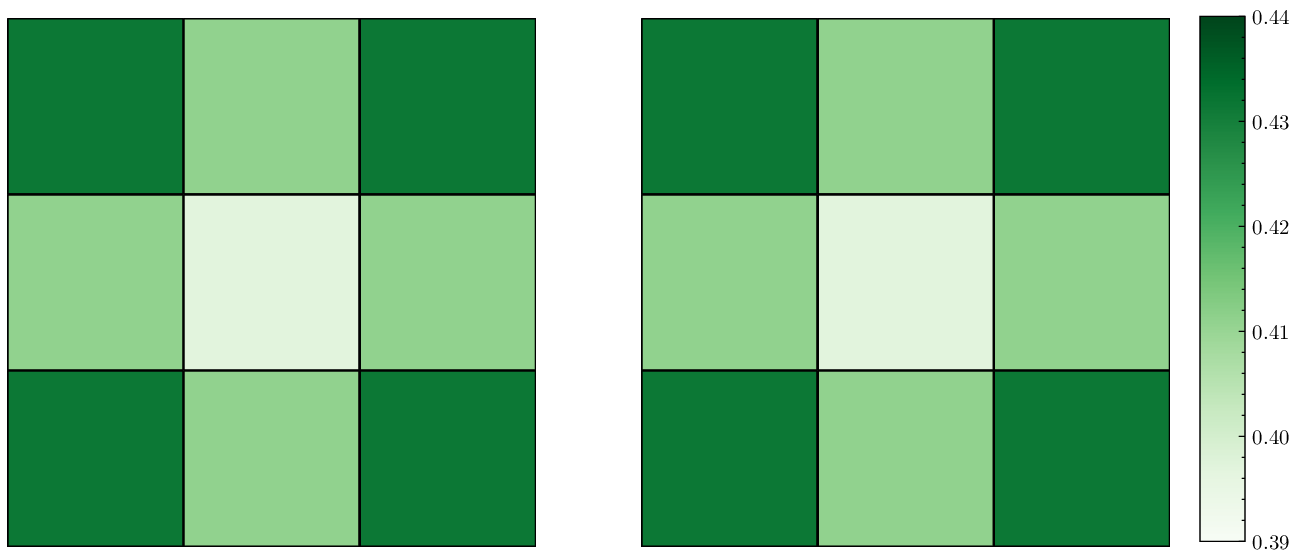}
        \label{fig:subfig2}
    }
    \caption{Skewness of the CSNS when using different matrix decomposition methods. \(\bm{\Sigma}\) is given by \((0.5 \, \text{diag}(\bm{W} \bm{1}_9 - \bm{W}) + 0.5 \bm{I}_9)^{-1}\). (Left) Skewness when data are arranged sequentially from the top left; (Right) Skewness when using square root decomposition with data order randomly shuffled.}
    \label{fig:skewness plot}
\end{figure}

If the symmetry does not hold under permutation, a change in the order of the data can affect the estimation results. To illustrate how the choice of the matrix decomposition method impacts symmetry, Figure \ref{fig:skewness plot} shows the skewness at each point on a \(3 \times 3\) grid for the model under different matrix decomposition methods. When using Cholesky decomposition, the skewness lacks symmetry, and reordering the data sequence causes random variations in skewness. In contrast, when using the matrix square root, the skewness exhibits symmetry and remains unchanged regardless of the order of points. This property holds not only for skewness but for the entire distribution. Consequently, using the matrix square root is generally preferable to decompose the variance matrix in cases where the ordering of data points is arbitrary.

Notably, when using the CAR model, spatial Markov properties are not preserved, although this is rarely a practical concern. Therefore, it is more appropriate to consider the CAR model merely as one of the design methods in constructing covariance matrices.

\subsection{Application to temporal modeling}
To incorporate the proposed model into a time series framework, first, we consider an \(\ell\)-order autoregressive (AR) model with the error term following the proposed distribution:
\begin{align}
\bm{s}_{\tau} &= \bm{A}_{1} (\bm{s}_{\tau-1} - \bm{\mu}_{\tau-1}) + \cdots + \bm{A}_\tau (\bm{s}_{\tau-\ell} - \bm{\mu}_{\tau-\ell}) + \bm{\xi}_{\tau}, \label{ARmodel}
\end{align}
where \(\tau > \ell\), \(\bm{A}_{1}, \ldots, \bm{A}_{\ell} \in \mathbb{R}^{N\times N}\), and \(\bm{s}_1 = \bm{\mu}_1 + \bm{\Sigma}^{1/2}\bm{t}_1, \ldots, \bm{s}_{\ell} = \bm{\mu}_{\ell} + \bm{\Sigma}^{1/2}\bm{t}_{\ell}\). Here, we assume that \(\bm{\xi}_{\tau}\) follows Equation \eqref{proposed model}.

The AR model is closed under the proposed model, and its mean and variance remain the same when assuming multivariate normal distribution is assumed for the error terms. Specifically, the model represented by Equation \eqref{ARmodel} is equivalent to the model expressed below:
\begin{align*}
\bm{s}_{1:T} = \bm{\mu}_{1:T} + \bm{R}^C (\bm{I}_T \otimes \bm{\Sigma}^{1/2}) \bm{t}_{1:T},
\end{align*}
where \(T\) is length of time; \(\otimes\) denotes the Kroneker product; \(\bm{s}_{1:T} = (\bm{s}_1^\top, \ldots, \bm{s}_T^\top)^\top\), \(\bm{t}_{1:T} = (\bm{t}_1^\top, \ldots, \bm{t}_T^\top)^\top\), and \(\bm{\mu}_{1:T} = (\bm{\mu}_1^\top, \ldots, \bm{\mu}_T^\top)^\top\). \(\bm{R}^C\) is the Cholesky decompositon of the covariance matrix of the AR model when \(\bm{\Sigma} = \bm{I}_K\). Each element of this submatrix is given as follows:
\begin{align*}
& (\bm{R}^C)_{\tau_1, \tau_1} = \bm{I}_N, & \text { for } \tau_1 = 1, \ldots, T, \\
& (\bm{R}^C)_{\tau_1, \tau_2} = \bm{O}_N, & \text { for } 1 \leq \tau_2 < \tau_1 \leq \tau, \\
& (\bm{R}^C)_{\tau_1, \tau_2} = \sum_{s=1}^{\tau} \bm{A}_{s} (\bm{R}^C)_{\tau_1-s, \tau_2}, & \text { for } \tau \leq \tau_2 < \tau_1 \leq T,
\end{align*}
where  \((\bm{R}^C)_{\tau_1, \tau_2}\) is the submatrix of \(\bm{R}^C\), consisting of rows \((\tau_1-1)N+1\) to \(\tau_1N\) and columns \((\tau_2-1)N+1\) to \(\tau_2N\), and \(\bm{O}_N\) is the \(N \times N\) zero matrix.

Unlike spatial models, the appearance of the Cholesky decomposition in the AR model can be explained by the fact that the order of data in the temporal domain carries significant meaning. This distinction can also be understood from the fact that spatial adjacency is represented by an undirected graph, whereas temporal adjacency is represented by a directed graph.

%% file: 5_bayesian_inference.tex
\section{Bayesian inference}
This section proposes a Bayesian estimation method for the parameters of the proposed model. Section 5.1 describes a Bayesian inference approach using the data augmentation method for general models, including spatial models. Since the data augmentation method is particularly effective for time series models, its application is detailed in Section 5.2.

\subsection{Bayesian inference for the proposed model}
To efficiently perform Bayesian inference for the posterior distribution of the proposed model, we propose a sampling method that combines data augmentation and Gibbs sampling. Firstly, \(\bm{s} \mid \bm{u}, \bm{\theta}, \bm{\mu}, \bm{\Sigma}\) follows a multivariate normal distribution:
\begin{align*}
    \mathcal{N}_N(\bm{\mu} + \bm{\Sigma}^{1/2}\bm{F}\bm{g}, \bm{\Sigma}^{1/2}\bm{F}\bm{H}\bm{H}\bm{F}(\bm{\Sigma}^{1/2})^\top),
\end{align*}
where \(\bm{F} = diag(f_1, \ldots, f_N)\), \(\bm{g} = (g_1, \ldots, g_N)^\top\), \(\bm{H} = diag(h_1, \ldots, h_N)\), with the elements depending on $\bm{u}$ and $\bm{\theta}$ (see Table \ref{tab:examples of the proposed model}). Therefore, \(\bm{s}\) conditioned on \(\bm{u}\) can be treated in the same manner as in models with a standard normal prior distribution. For example, given the observation model  
\begin{align*}
    \bm{y} = \bm{s} + \bm{\varepsilon}, \quad \bm{\varepsilon} \sim \mathcal{N}_N(\bm{0}_N, \sigma^2\bm{I}_N),
\end{align*}
the posterior distribution of \(\bm{s} \mid \bm{y}, \bm{u}, \bm{\theta}, \bm{\mu}, \bm{\Sigma}\) can be expressed analytically as follows:
\begin{align*}
    \mathcal{N}_N(\bm{\mu}_{\bm{s}} + \bm{\Sigma}_{\bm{s}}(\bm{\Sigma}_{\bm{s}}+\sigma^2\bm{I}_N)^{-1}(\bm{y} - \bm{\mu}_{\bm{s}}), \bm{\Sigma}_{\bm{s}} + \bm{\Sigma}_{\bm{s}}(\bm{\Sigma}_{\bm{s}}+\sigma^2\bm{I}_N)^{-1}\bm{\Sigma}_{\bm{s}}),
\end{align*}
where \(\bm{\mu}_{\bm{s}} = \bm{\mu} + \bm{\Sigma}^{1/2}\bm{F}\bm{g}\) and \(\bm{\Sigma}_{\bm{s}} = \bm{\Sigma}^{1/2}\bm{F}\bm{H}\bm{H}\bm{F}(\bm{\Sigma}^{1/2})^\top\). Similarly, if the prior for \(\bm{\mu}\) is a multivariate normal distribution, then \(\bm{\mu} \mid \bm{s}, \bm{u}, \bm{\theta}, \bm{\Sigma} \) also follows a multivariate normal distribution. This property is useful for deriving coefficients in linear regression and other related models. The variables \(\bm{u}, \bm{\theta}\) influences \(\bm{s}\) only through \(\bm{t} = \bm{\Sigma}^{-1/2}(\bm{s} - \bm{\mu})\). Using this property, the elements of \(\bm{u}, \bm{\theta}\) may be sampled individually from \(u_i, \theta_i \mid t_i\) instead of \(\bm{u}, \bm{\theta} \mid \bm{s}, \bm{\mu}, \bm{\Sigma}\) after sampling $\bm{s}$.
Therefore, the sampling from the posterior distribution \(u_i \mid t_i, \theta_i\) and \(\theta_i \mid t_i, u_i\) can be implemented in parallel. Additionally, in some distributions, such as the t, Laplace, CSN distributions, \(p(u_i \mid t_i, \theta_i)\) is represented analytically. For example, in a CSNS distribution, 
\begin{align*}
u_i \mid t_i, \theta_i \sim \mathcal{TN}\left(\frac{\theta_i}{1+\theta_i^2}\left(\sqrt{\frac{(\pi-2)\theta_i^2+\pi}{\pi}} t_i + \sqrt{\frac{2}{\pi}}\theta_i\right), \frac{1}{1+\theta_i^2}\right).
\end{align*}
This method significantly reduces computational costs. Sampling from an \(N\)-dimensional CSN distribution requires a computational complexity of \(O(N^3)\) when sampling from an \(N\)-dimensional multivariate truncated normal distribution \citep{botev2017normal}. In contrast, the sampling method described above only requires sampling \(u_i\) from independently distributed truncated normal distributions, resulting in a computational complexity of \(O(N)\).

\subsection{Bayesian inference for temporal model}
As the AR model is closed under the proposed framework, the aforementioned data augmentation method can also be applied to the AR model. Furthermore, because the transitions of \(\bm{s}_{1:T}\) form a linear Gaussian state-space model, a more efficient method known as forward filtering backward sampling (FFBS) method can be applied \citep{carter1994gibbs, fruhwirth1994data}. For simplicity, we consider only the first-order AR here; however, the following techniques can generally be applied to higher-order models as well. The model with observation noise is expresed as follows:
\begin{align}
\bm{y}_{\tau} &= \bm{s}_{\tau} + \bm{\varepsilon}_{\tau}, \\
\bm{s}_{\tau} &= \bm{A}_{1} (\bm{s}_{\tau-1} - \bm{\mu}_{\tau-1}) + \bm{\xi}_{\tau},  \notag \\
\bm{\varepsilon}_{\tau} &\sim \mathcal{N}_N(\bm{0}_N, \sigma^2 \bm{I}_N) \notag \\
\bm{\xi}_{\tau} &= \bm{\mu}_{\tau} + \bm{\Sigma}^{1/2} \bm{t}_{\tau}. \notag
\end{align}

Regarding the posterior distribution of \(\bm{s}_{1:T}\), conditioned on \(\bm{u}_{1:T}\), it becomes a normal distribution allowing the use of the FFBS method. Let \(\bm{\mu}_{\tau_1|\tau_2}\) and \(\bm{\Sigma}_{\tau_1|\tau_2}\) be the mean and the variance of \(\bm{s}_{\tau_1} | \bm{y}_{1:\tau_2}, \bm{u}_{1:\tau_2}, \bm{\theta}_{1:\tau_2}, \bm{\mu}_{1:\tau_2}, \bm{\Sigma}, \bm{A}_1, \sigma\), then
\begin{align*}
\bm{\mu}_{\tau\mid\tau-1} &= \bm{A}_1 \bm{\mu}_{t-1|t-1} + \bm{\mu}_t + \bm{\Sigma}^{1/2}\bm{F}\bm{g}, \\
\bm{\Sigma}_{\tau\mid\tau-1} &= \bm{A}_1 \bm{\Sigma}_{t-1|t-1} \bm{A}_1^\top + \bm{\Sigma}^{1/2}\bm{F}\bm{H}\bm{H}\bm{F}(\bm{\Sigma}^{1/2})^\top, \\
\bm{\mu}_{\tau\mid\tau} &= \bm{\mu}_{\tau\mid\tau-1} + \frac{1}{\sigma^2}\bm{\Sigma}_{\tau\mid\tau}(\bm{y}_t - \bm{\mu}_{\tau\mid\tau-1}), \\
\bm{\Sigma}_{\tau\mid\tau} &= \sigma^2(\bm{\Sigma}_{\tau\mid\tau-1} + \sigma^2 \bm{I}_K)^{-1}\bm{\Sigma}_{\tau\mid\tau-1},
\end{align*}
where \(\bm{\mu}_{1|0}\) is \(\bm{\mu}_1 + \bm{\Sigma}^{1/2}\bm{F}\bm{g}\), and variance is \(\bm{\Sigma}_{1|0} = \bm{\Sigma}^{1/2}\bm{F}\bm{H}\bm{H}\bm{F}(\bm{\Sigma}^{1/2})^\top\). By using the above formulation iteratively, the distribution of \(\bm{s}_{\tau} | \bm{y}_{1:\tau}, \bm{u}_{1:\tau}, \bm{\theta}_{1:\tau}, \bm{\mu}_{1:\tau}, \bm{\Sigma}, \bm{A}_1, \sigma\) can be calculated for \(\tau = 1 , \ldots, T\). After calculation, \(\bm{s}_{\tau} | \bm{y}_{1:T}, \bm{u}_{1:T}, \bm{\theta}_{1:T}, \bm{\mu}_{1:T}, \bm{\Sigma}, \bm{A}_1, \sigma\) can be sequentially backward sampled:
\begin{align*}
\bm{\mu}_{\tau|T} &= \bm{\mu}_{\tau|t} + \bm{A}_1 \bm{\Sigma}_{\tau\mid\tau} \bm{\Sigma}_{\tau+1|\tau}^{-1}(\bm{s}_{\tau+1|T}^{(i)} - \bm{\mu}_{\tau+1|\tau}), \\    
\bm{\Sigma}_{\tau|T} &= \bm{\Sigma}_{\tau|\tau} - \bm{\Sigma}_{\tau|\tau} \bm{\Sigma}_{\tau+1|\tau}^{-1} \bm{\Sigma}_{\tau|\tau},
\end{align*}
where \(\bm{s}_{\tau+1|T}^{(i)}\) is a sample of \(\bm{s}_{\tau+1} | \bm{y}_{1:T}, \bm{u}_{1:T}, \bm{\theta}_{1:T}, \bm{\mu}_{1:T}, \bm{\Sigma}, \bm{A}_1, \sigma\). 

%% file: 6_simulation_studies.tex
\section{Simulation studies}

This study investigated the inverse estimation performance for spatio-temporal data through simulations. Specifically, using the case of the CSNS distribution as an example, we examined whether the estimation performance improves by properly modeling heterogeneity in higher-order moments. To this end, four types of heterogeneity were investigated. The models are defined as follows: the model with \(\theta_{tk} = 0\) for \(t=1,\ldots,T, k=1,\ldots,K\) is referred to as the zero theta (ZT) model, corresponding to the Gaussian regression model; the model with \(\theta_{tk} = \theta\) for \(t=1,\ldots,T, k=1,\ldots,K\) is referred to as the fixed theta (FT) model; the model with \(\theta_{t1} = \cdots = \theta_{tK} = \theta_t\)  for \(t=1,\ldots,T\) is referred to as the temporal varying theta (TT) model; and the model with \(\theta_{1k} = \cdots = \theta_{Tk}\) for \(k=1,\ldots,K\) is referred to as the spatial varying theta (ST) model.

The observation model is as follows:
\begin{align*}
\bm{y}_{\tau} &= \beta_1 \bm{x}_{\tau} + \beta_2 \bm{1}_K + \bm{s}_{\tau} + \bm{\varepsilon}_{\tau}, \notag \\
\bm{s}_{\tau} &= \rho \bm{s}_{\tau-1} + \bm{\xi}_{\tau}, \\
\bm{\varepsilon}_{\tau} &\sim \mathcal{N}_N(\bm{0}_N, \sigma^2 \bm{I}_N) \notag
\end{align*}
where \(\bm{\xi}_{\tau}\) follows a CSNS distribution, depending on the \(\theta_{tk}\) parameter, using the matrix square root. The covariance matrix \(\bm{\Sigma}\), which models spatial correlation, is assumed to follow Equation \eqref{Leroux model}. The time length \(T\) for parameter estimation was set to \(50\), whereas the time length for evaluating prediction accuracy, \(T_{future}\), was set to \(2\). The location is represented by a \(5 \times 5\) grid. The adjacency matrix is defined by assigning a value of \(1\) to the neighboring pairs that share a border with separating grids, and \(0\) to all others. The features \(\bm{x}_{\tau}\) is generated as a random sample from the standard normal distribution.

The true values for the simulation model for each case are as follows: in Case 1, we used the ZT model; in Case 2, the FT model with \(\theta=2.5\); in Case 3, the TT model with \(\theta_{\tau} \sim N(2.5, 3^2), t = 1, \ldots, T\), where \(\theta_{\tau}\) were generated as random variables; and in Case 4, the ST model with \(\theta_k \sim N(2.5, 3^2), k = 1, \ldots, K\), where \(\theta_k\) is generated as random variables. For the remaining parameters, the coefficient \(\beta_1\) for randomly generated features was set to \(1.0\), intercept \(\beta_2\) to \(0.5\), standard deviation of the observation noise \(\sigma\) to \(0.1\), coefficient of the temporal AR model \(\rho\) to \(0.5\), parameter \(\omega\), which determines the variability of \(\bm{\theta}\), to \(1.0\), and parameter \(\eta\), which controls the strength of spatial correlation, to \(0.5\).

The prior distributions for the estimation model were shared across all four models, following \cite{lee2018spatio}. We employ a hierarchical distribution as the prior distributions for \(\bm{\theta}_{1:T}\). The priors are as follows:
\begin{align*}
\theta_i &\sim \mathcal{N}_1(\mu_\theta, \nu_\theta^2), \\
\mu_\theta &\sim \mathcal{N}_1(0, 100 \nu_\theta^2), \\
\nu_\theta^2 &\sim \mathcal{IG}(1, 0.01) \\
\bm{\beta} &\sim \mathcal{N}_p(\bm{0}_p, 100 \bm{I}_p), \\
\sigma^2 &\sim \mathcal{IG}(1, 0.01), \\
\rho &\sim \mathcal{U}(0, 1), \\
\omega^2 &\sim \mathcal{IG}(1, 0.01), \\
\eta &\sim \mathcal{U}(0, 1),
\end{align*}
where \(\mathcal{U}(0, 1)\) denotes the uniform distribution from \(0\) to \(1\).

\begin{figure}[!hbt]
  \centering
  \includegraphics[width=0.95\linewidth]{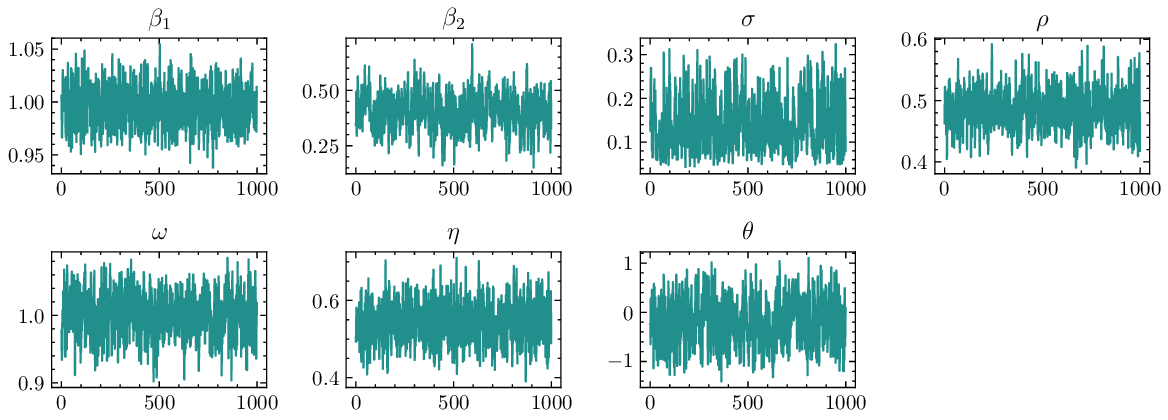}
  \caption{Trace plots of samples from posterior distribution.}
  \label{fig:trace plots}
\end{figure}

Using the sampling method described in Section 4.3, we obtained 1,000 samples from the posterior distribution for each simulation to evaluate estimation performance. During the sampling process, more than 50,000 samples were discarded corresponding to the burn-in period based on the results of trace plots, and thinning was applied to ensure that the autocorrelation was below 0.3. Figure \ref{fig:trace plots} presents an example of a trace plot for several parameters of the FT model in Case 1, after discarding burn-in samples and applying thinning. The plot confirms that convergence was sufficient.

\begin{figure}[!hbt]
  \centering
  \includegraphics[width=0.90\linewidth]{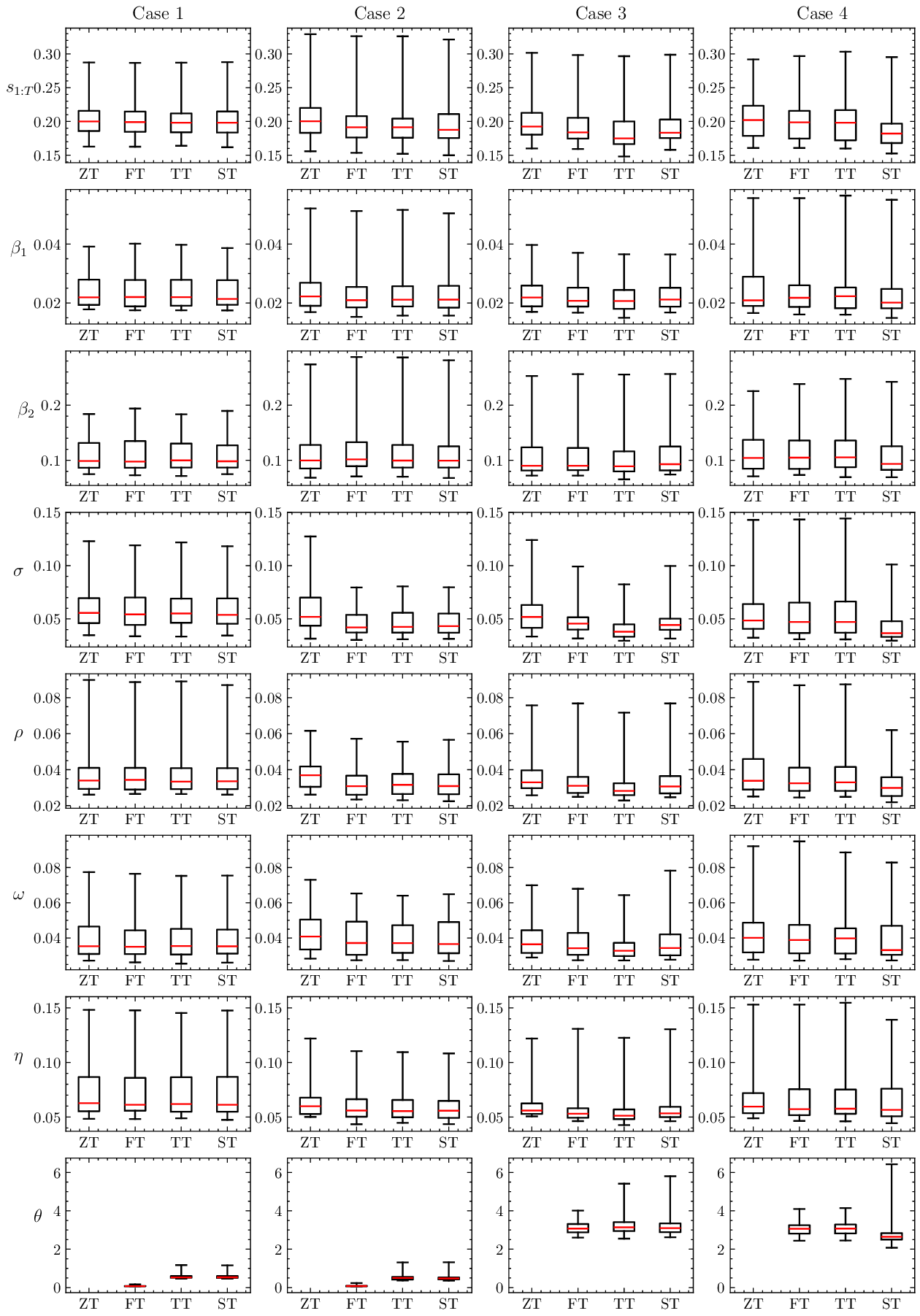}
  \caption{RMSE in each case for each model.}
  \label{fig:rmse}
\end{figure}

Figure \ref{fig:rmse} presents the root mean squared errors (RMSE) of each parameter for the four cases and four estimation models that were simulated 50 times. Box plots are used to illustrate each case. In the ZT model, \(\bm{\theta}\) was fixed to \(\bm{0}_{TK}\) and not estimated; therefore, the RMSE plot for \(\bm{\theta}\) is left blank for the ZT model.

First, for Case 1, assuming normally distributed latent variables, the estimation performance is nearly identical across all four models. This indicates that even when the true model follows a normal distribution, accounting for higher-order moments does not degrade estimation performance. For Case 2, the estimation performance of the ZT model is generally poor; however no significant difference in performance is observed for the other three models. This suggests that when \(\theta\) is fixed to a single value, consideration of heterogeneity in higher-order moments does not impair estimation performance. Finally, in Cases 3 and 4, the true models (TT and ST) demonstrate improved estimation performance for most parameters. Notably, the improvements are significant for the standard deviation of observational noise (\(\sigma\)), time-series coefficient (\(\rho\)), and variability of random effects (\(\omega\)). These results suggest that modeling heterogeneity in higher-order moments enhances parameter estimation performance. For \(\theta\), the FT model performed well in Cases 1 and 2, whereas in Case 3, the variation is slightly different with no notable differences among the three models. In Case 4, despite increased variation, the ST model improved estimation performance.

\begin{figure}[!hbt]
  \centering
  \includegraphics[width=0.95\linewidth]{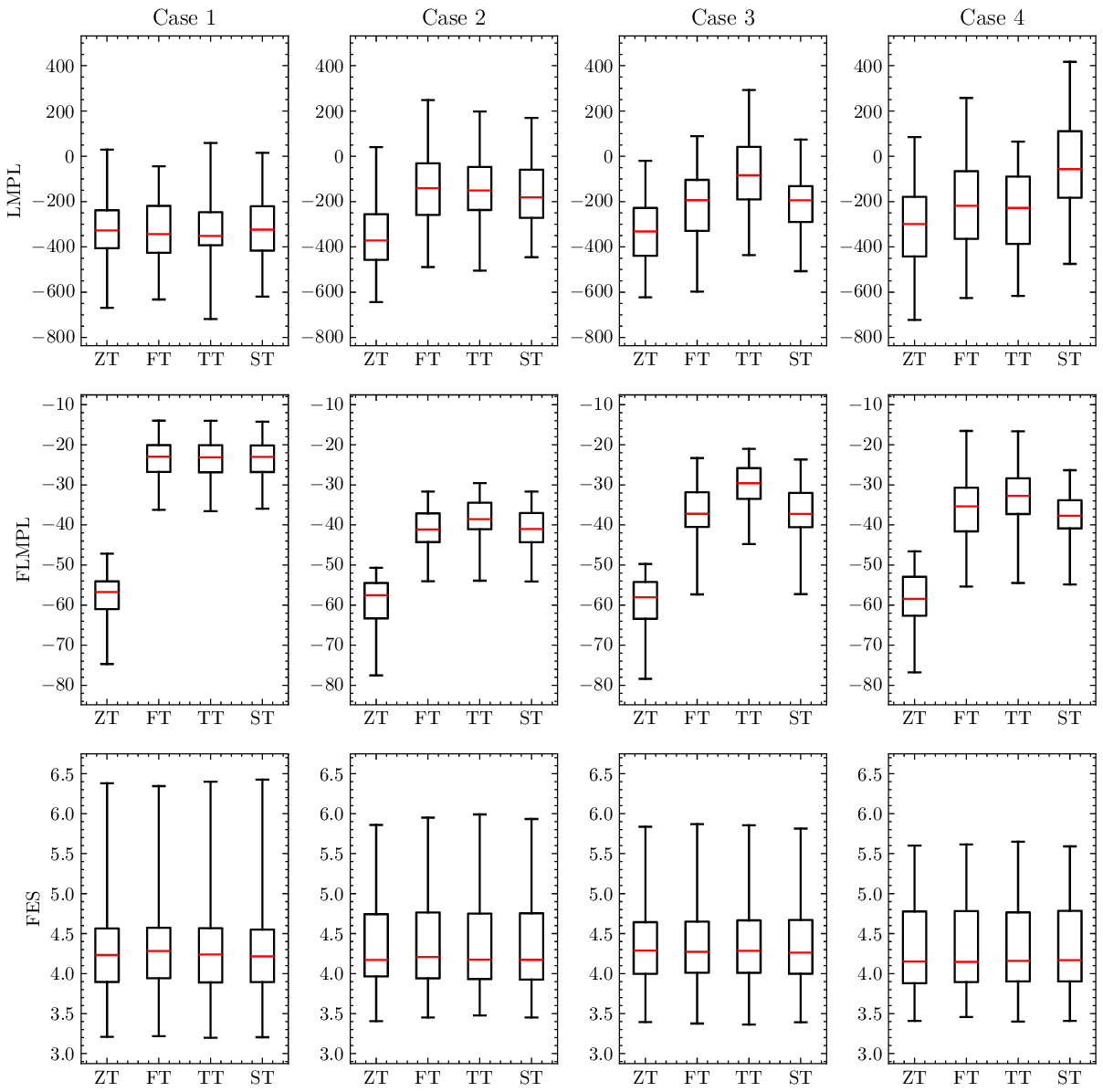}
  \caption{Predictive accuracy in each case for each model.}
  \label{fig:predictive_accuracy}
\end{figure}

Similarly, Figure \ref{fig:predictive_accuracy} displays predictive accuracy, which was evaluated using log marginal predictive likelihood (LMPL), LMPL in the future data (FLMPL), and RMSE of \(y\) in the future data (FRMSE). The calculation method for LMPL followed that used in the CarBayesST package \citep{lee2018spatio}. FLMPL is an estimator of LMPL for the future data \(y_{T+1:T+T_{future}}\) conditioned on the learning data \(y_{1:T}\), specifically \(\log p(y_{T:T+T_{future}}|y_{1:T})\). FLMPL is calculated from samples of the posterior distribution as follows:
\begin{align*}
\mathrm{FLMPL} = \frac{1}{S} \sum_{i=1}^S \log p(\bm{y}_{T+1:T+T_{future}} | \bm{s}^{(i)}, \bm{\theta}^{(i)}, \bm{\beta}^{(i)}, \sigma^{(i)}, \rho^{(i)}, \omega^{(i)}, \eta^{(i)}),
\end{align*}
where \(\bm{s}^{(i)}, \bm{\theta}^{(i)}, \bm{\beta}^{(i)}, \sigma^{(i)}, \rho^{(i)}, \omega^{(i)}, \eta^{(i)}\) is a sample of parameters from the posterior distribution conditioned on \(\bm{y}_{1:T}\), and \(S\) is the number of samples. Similarly, FRMSE is the RMSE of \(\bm{y}_{T+1:T+T_{future}}\) and is numerically calculated as follows:
\begin{align*}
\mathrm{FRMSE} &= \sqrt{\frac{1}{S} \sum_{i=1}^S || \bm{y}_{T+1:T+T_{future}} - \hat{\bm{y}}_{T+1:T+T_{future}}^{(i)}||^2}.
\end{align*}
where \(\hat{\bm{y}}_{T+1:T+T_{future}}^{(i)}\) is a sample drawn from \(p(\bm{y}_{T+1:T+T_{future}} | \bm{y}_{1:T})\), and \(|| \cdot ||\) is the Euclidean norm.

For LMPL, in Case 1, there is no significant difference among the four models, whereas in Cases 2, 3, and 4, the true model shows higher values. This indicates that selecting the true model improves the fit to the observed data. In contrast, for FLMPL, the ZT model performs poorly across all cases, with the TT model consistently performing slightly better than the FT and ST models. In this scenario, the standard deviation of the observational noise is 0.1, whereas the FRMSE ranges from approximately 1.1 to 1.5. This can be attributed to the assumption that \(\theta\) varies over time, causing the model to exhibit heavier tails, which influences prediction. Finally, for FRMSE, there is little variation across the cases. This can be interpreted as the CSNS distribution maintaining the mean and variance, resulting in minimal differences in squared errors of the observations.

%% file: 7_application.tex
\section{Application}

\begin{figure}[!hbt]
  \centering
  \includegraphics[width=\linewidth]{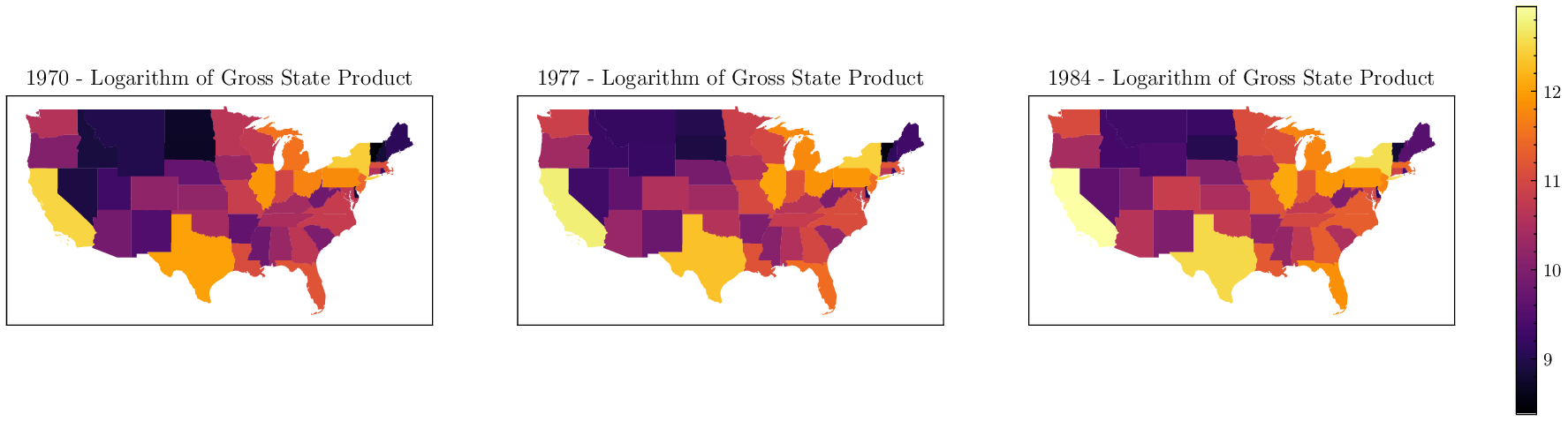}
  \caption{Logarithm of gross state product in 1970, 1977, and 1984.}
  \label{fig:plot of y}
\end{figure}

This study focuses on identifying the production functions for 48 states in the United States, using annual data from 1970 to 1986, following the model outlined in \cite{badi2021econometric}. A production function represents how capital and labor affect output. We assumed a Cobb-Douglas type production function, where output is expressed as the product of capital and labor raised to respective exponents. By taking the logarithm of the Cobb-Douglas production function and rewriting it in a form that allows for coefficient estimation from observed data, it can be expressd as follows:
\begin{align*}
y_{tk} &= \beta_{intercept} + \beta_{\mathrm{KP}} \log \mathrm{KP}_{tk} + \beta_{\mathrm{KH}} \log \mathrm{KH}_{tk} + \beta_{\mathrm{KW}} \log \mathrm{KW}_{tk} \notag \\
&\qquad + \beta_{\mathrm{KO}} \log \mathrm{KO}_{tk} + \beta_{\mathrm{L}} \log \mathrm{L}_{tk} + \beta_{\mathrm{Unemp}} \mathrm{Unemp}_{tk} + s_{tk} + \epsilon_{tk},
\end{align*}
Here, \(y\) represents the logarithm of the gross state product, with the features including private capital stock (\(\mathrm{KP}\)), highways and streets (\(\mathrm{KH}\)), water and sewer facilities (\(\mathrm{KW}\)), other public buildings and structures (\(\mathrm{KO}\)), nonagricultural payroll employment (\(\mathrm{L}\)), and the state unemployment rate (\(\mathrm{Unemp}\)). All features, except the unemployment rate, are expressed in natural logarithms. The data spanning from 1970 to 1984 were utilized as the training dataset, whereas the data from 1985 to 1986 were used to validate the predictive accuracy of the models. Figure \ref{fig:plot of y} displays the value of \(\bm{y}\), the logarithm of the gross state product for each state. Data from the first (1970), middle (1977), and final year (1984) are plotted. 

For the models, we compared the estimation results of ZT, FT, TT, and ST, using the same approach and priors as that of the simulation study. We set sufficiently long burn-in and thinning periods and obtained \(4000\) samples from the posterior distribution to evaluate the estimation results.

Table \ref{tab:estimation_results} displays the median and 90\% credible interval of samples drawn from the posterior distributions of parameters for each model. The medians of the posterior distributions of the all parameters exhibit similar patterns, although a slight variation is observed in the estimates of the regression coefficients. This variation can be attributed to the  proposed model, which preserves both the mean and variance.

\begin{table}[!hbt]
\centering
\caption{Estimation results of parameters-50\% percentile and 90\% credible interval}
\label{tab:estimation_results}

\begin{minipage}{\linewidth}
\centering
\begin{tabular}{lcc}
\hline
 & ZT & FT \\ \hline
\(\beta_{intercept}\) & 1.197 (1.033--1.322) & 1.317 (1.144--1.451) \\
\(\beta_{\mathrm{KP}}\) & 0.404 (0.371--0.429) & 0.386 (0.352--0.413) \\
\(\beta_{\mathrm{KH}}\) & 0.143 (0.101--0.175) & 0.120 (0.079--0.152)  \\
\(\beta_{\mathrm{KW}}\) & 0.080 (0.054--0.099) & 0.080 (0.054--0.101) \\
\(\beta_{\mathrm{KO}}\) & -0.007 (-0.037--0.016) & -0.018 (-0.048--0.007) \\
\(\beta_{\mathrm{L}}\) & 0.470 (0.431--0.501) & 0.513 (0.468--0.547)  \\
\(\beta_{\mathrm{Unemp}}\) & -0.007 (-0.010-- -0.005) & -0.006 (-0.009-- -0.004)  \\
\(\sigma\) & 0.015 (0.014--0.016) & 0.015 (0.014--0.016)  \\
\(\rho\) & 0.953 (0.929--0.973) & 0.953 (0.927--0.972) \\
\(\omega\) & 0.050 (0.047--0.052) & 0.049 (0.046--0.052)\\
\(\eta\) & 0.756 (0.640--0.830) & 0.742 (0.627--0.822)\\\hline
\end{tabular}
\end{minipage}

\vspace{1em}

\begin{minipage}{\linewidth}
\centering
\begin{tabular}{lcccccc}
\hline
 & TT & ST \\ \hline
\(\beta_{intercept}\)& 1.431 (1.253--1.566) & 1.378 (1.211--1.522) \\
\(\beta_{\mathrm{KP}}\) & 0.366 (0.331--0.394) & 0.328 (0.293--0.356) \\
\(\beta_{\mathrm{KH}}\) & 0.114 (0.072--0.148) & 0.179 (0.134--0.212) \\
\(\beta_{\mathrm{KW}}\) & 0.081 (0.054--0.102) & 0.081 (0.055--0.102) \\
\(\beta_{\mathrm{KO}}\) & -0.025 (-0.056-- -0.001) & -0.021 (-0.053--0.004) \\
\(\beta_{\mathrm{L}}\) & 0.538 (0.495--0.572) & 0.527 (0.486--0.560) \\
\(\beta_{\mathrm{Unemp}}\) & -0.004 (-0.007-- -0.002) & -0.007 (-0.010-- -0.005) \\
\(\sigma\) & 0.015 (0.014--0.016) & 0.014 (0.013--0.015) \\
\(\rho\) & 0.947 (0.924--0.964) & 0.960 (0.939--0.975) \\
\(\omega\) & 0.047 (0.044--0.049) & 0.043 (0.041--0.046) \\
\(\eta\) & 0.569 (0.447--0.662) & 0.697 (0.565--0.785) \\
\hline
\end{tabular}
\end{minipage}

\end{table}

\begin{figure}[!hbt]
  \centering
  \includegraphics[width=\linewidth]{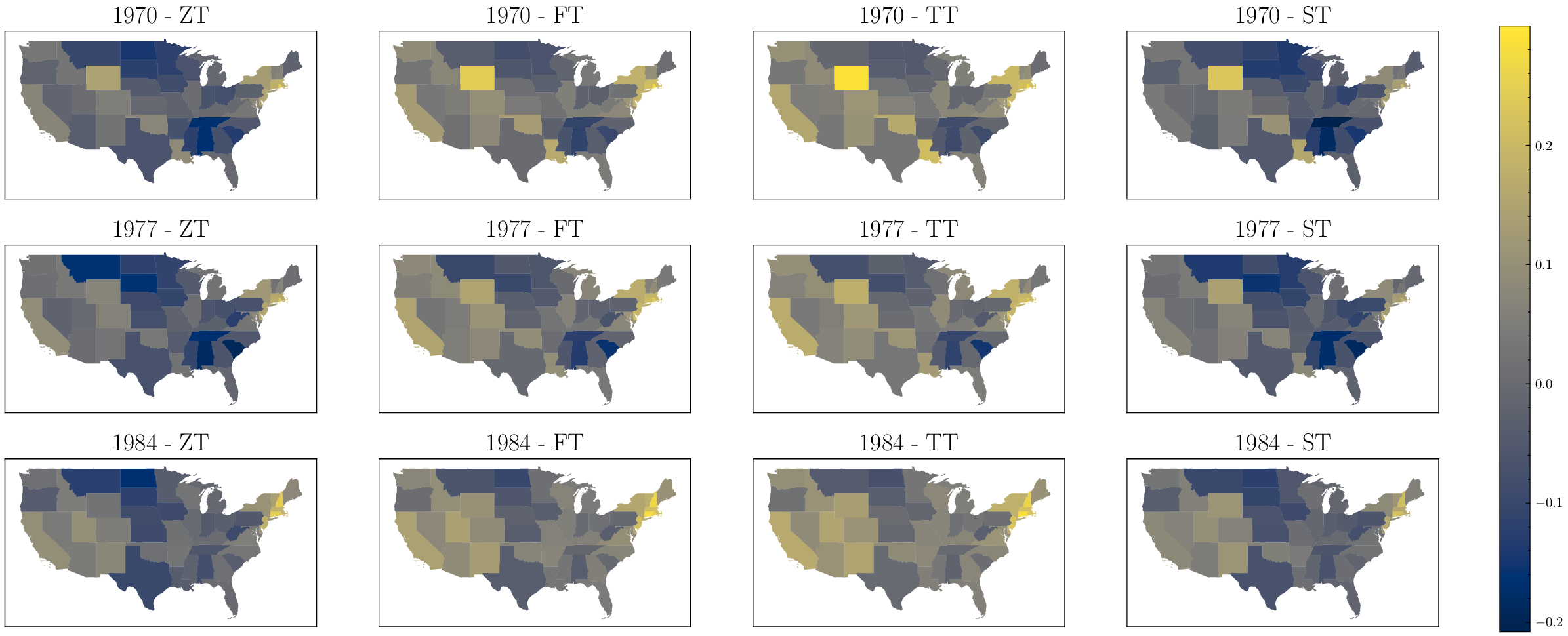}
  \caption{Median of \(\bm{s}\) for each model}
  \label{fig:median_of_application}
\end{figure}

Figure \ref{fig:median_of_application} displays the median of the posterior distribution of \(\bm{s}_{\tau}\) for each state in 1970, 1977, and 1984 under each model. The figure reveals that, while the trends in the estimates of \(\bm{s}_{\tau}\) are similar across the four models, the actual values differ significantly. For example, the posterior median of \(s\) for Wyoming in 1970 under the ZT model is approximately \(0.15\), whereas it is \(0.28\) under the TT model. Certain states appear in much brighter colors under the CSNS model. For instance, Wyoming around 1970 is depicted in a brighter color, indicating higher productivity than expected based on its capital and infrastructure. In 1970, Wyoming was the second least populous state \citep{united1973census}, exhibiting low feature values, with \(L\) being the smallest. However, owing to its abundant resources such as coal, crude oil, and natural gas \citep{us2023production}, Wyoming demonstrated a total production that surpassed predictions based on its features. The states appearing in brighter colors under the CSNS model suggest that this model assigns greater probability to positive values than the normal model, because \(\theta_i\) is positive. Additionally, in the ST model, differences in shading intensity are more distinct separated than in the FT and TT models.

\begin{figure}[!hbt]
  \centering
  \includegraphics[width=\linewidth]{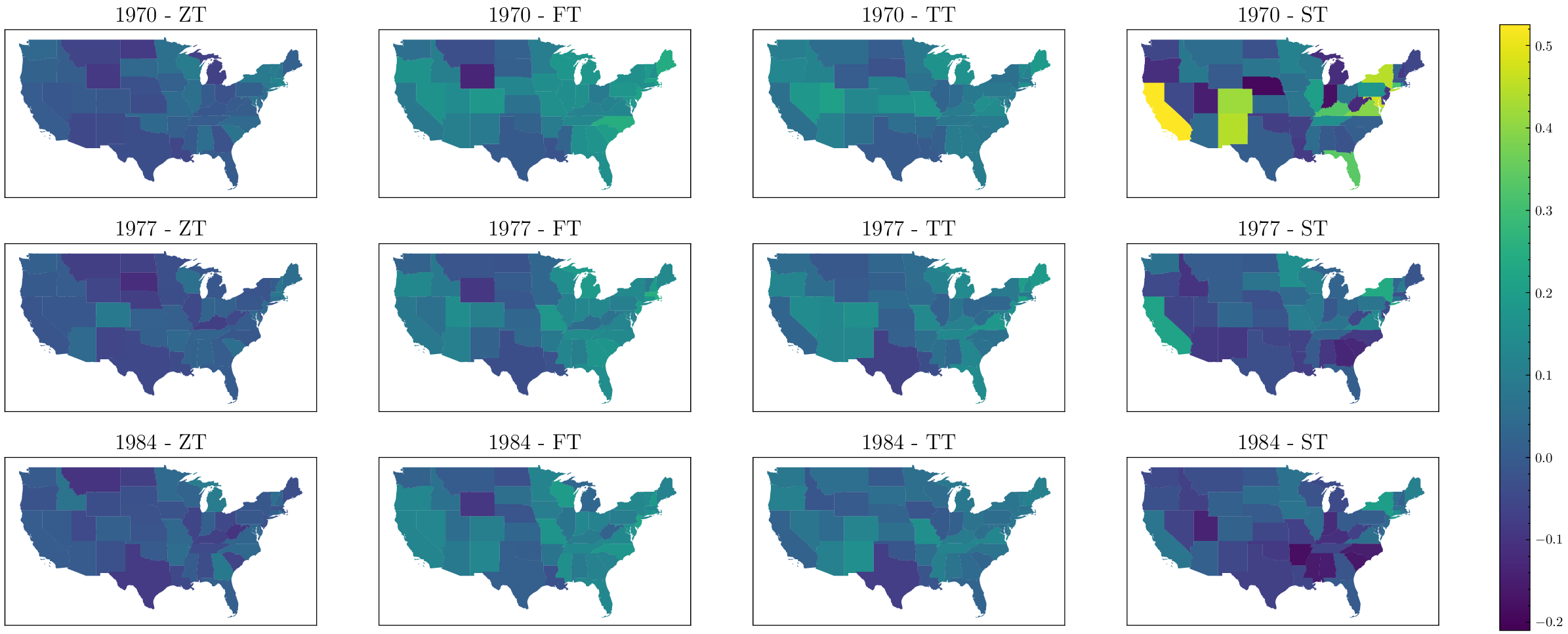}
  \caption{Skewness of \(\bm{s}\) for each model.}
  \label{fig:skewness_of_application}
\end{figure}

Figure \ref{fig:skewness_of_application} depicts the skewness of the posterior distributions in each model. In this figure, in the ST model for 1970, the differences in skewness are more distinct than in the other models. This is attributed to the ST model’s ability to capture the spatial heterogeneity of higher-order moments. In 1977 and 1984, these differences in skewness are less pronounced, presumably because the variance increases over time, making it more difficult for the skewness to become substantially larger.

Finally, Table \ref{tab:prediction accuracy in produc} presents the evaluation of predictive accuracy. Compared with the ZT model, the FT, TT and ST model show improvements in all evaluated metrics. Even though \(\bm{y}\) has undergone a logarithmic transformation, considering the heterogeneity of higher-order moments leads to enhanced estimation performance. In particular, the TT and ST models outperform the FT model, suggesting that incorporating both temporal and spatial heterogeneity in higher-order moments improves model fit and predictive accuracy.

\begin{table}[!hbt]
\centering
\begin{tabular}{lcccc}
\hline
& LMPL & FLMPL & FRMSE\((\times10^{-2})\)  \\ \hline
ZT & 1908.433 & 203.602 & 5.945 \\
FT & 1920.203 & 254.741 & 5.875 \\
TT & \textbf{1947.617} & \textbf{268.959} & 5.680 \\
ST & 1945.871 & 255.556 & \textbf{5.302} \\ \hline
\end{tabular}
\caption{Prediction accuracy in application.}
\label{tab:prediction accuracy in produc}
\end{table}

%% file: 8_conclusion.tex
\section{Conclusion}

In this study, we propose a method that enables the treatment of the heterogeneity of the higher order moments while preserving the mean and variance, crucial for maintaining parameter interpretability (see Section 1). After demonstrating closure under linear transformations of the proposed model, we analytically derived its skewness and kurtosis. In addition, applying it to spatio-temporal data, we examined how the properties of the model change depending on the chosen matrix decomposition method. In particular, using the matrix square root preserves symmetry with respect to ordering, making it suitable for data with inherent symmetry. For time-series data, we showed that the Cholesky decomposition corresponds to the AR model representation. We then proposed an efficient Bayesian inference method employing data augmentation and the FFBS algorithm.

In the simulation study, we compared the estimation performance of the four models that incorporated heterogeneity in higher-order moments. The results demonstrated that appropriately accounting for higher-order moment heterogeneity improves parameter estimation performance, model fit, and predictive accuracy.

For a real-data application, we applied the CSNS model to identify production functions across the United States. Although parameters other than \(\bm{s}\) generally exhibited similar trends, the posterior distribution of \(\bm{s}\) revealed that skewness varies substantially by location, particularly when spatial heterogeneity in higher-order moments is considered. Furthermore, both the TT and ST models showed improvements in terms of model fit and predictive performance metrics, indicating that even when the response variable is log-transformed, accounting for higher-order moment heterogeneity enhances estimation performance.

An open question for future research is the development of a model that simultaneously preserves the mean and variance, maintains spatial conditional independence, and accommodates heterogeneity in higher-order moments. Furthermore, as the proposed model encompasses a broad class of models, a potential extension may explore more complex structures where each dimension follows a different distribution, which remains unaddressed in this study.

The Julia implementation of CSNS model is available on GitHub: \\(https://github.com/Kuno3/CsnSubclass).